\documentclass[12pt,draftclsnofoot,onecolumn]{IEEEtran}
\usepackage{amsfonts,subfigure,multicol,color,verbatim,
graphicx,cite,epsfig,amssymb,amsmath,cases,bm,algorithm,
algorithmic,xcolor,multirow,url,stfloats}

\begin{document}
\newcommand{\loflabel}{Figure}
\newcommand{\lotlabel}{Table}

%
%


\title{Network Slicing in Fog Radio Access Networks: Issues and Challenges
\thanks{Hongyu Xiang (e-mail: xianghongyu88@163.com), Wenan Zhou (e-mail: zhouwa@bupt.edu.cn), and Mugen Peng (e-mail:
pmg@bupt.edu.cn) are with the Key Laboratory of Universal Wireless Communications (Ministry of Education), Beijing
University of Posts and Telecommunications, Beijing, China. Mahmoud Daneshmand (e-mail: mdaneshm@stevens.edu) is with
the Stevens Institute of Technology, Hoboken, NJ.}}
\author{\IEEEauthorblockN{Hongyu Xiang, Wenan Zhou, Mahmoud Daneshmand, and Mugen Peng}}

\date{\today}
\maketitle

\begin{abstract}
Network slicing has been advocated by both academia and industry as a cost-efficient way to enable
operators to provide networks on an as-a-service basis and meet the wide range of use cases that the
fifth generation wireless network will serve. The existing works on network slicing are mainly targeted
at the partition of the core network, and the prospect of network slicing in radio access networks
should be jointly exploited. To solve this challenge, an enhanced network slicing in fog radio access
networks (F-RANs), termed as access slicing, is proposed. This article comprehensively presents a
novel architecture and related key techniques for access slicing in F-RANs. The proposed hierarchical
architecture of access slicing consists of centralized orchestration layer and slice instance layer, which
makes the access slicing adaptively implement in an convenient way. Meanwhile, key techniques and
their corresponding solutions, including the radio and cache resource management, as well as the social-aware
slicing, are presented. Open issues in terms of standardization developments and field trials are
identified.
\end{abstract}


\newpage

\section{Introduction}


Driven by the emerging applications of mobile Internet and Internet of Things (IoTs), the fifth
generation (5G) wireless communication systems are expected to satisfy diverse use cases and
business models\textcolor[rgb]{1.00,0.00,0.00}{\cite{0}}.
Unfortunately, the legacy cellular network architectures are originally designed for mobile broadband
consumers, without considering the characteristics of emerging massive machine-type communications (mMTC) and ultra-reliable MTC (uMTC)\textcolor[rgb]{1.00,0.00,0.00}{\cite{1}}.
To meet the diverse
use cases and business models in 5G, network slicing is proposed recently to flexibly provide a
soft-defined networking in a cost-efficient way.


In the concept of network slicing\textcolor[rgb]{1.00,0.00,0.00}{\cite{2}}, the network entities are sliced into multiple isolated
network slice instances with appropriate network functions and
radio access technologies. Network slicing has attracted a lot of
interests from both academia and industry.
In\textcolor[rgb]{1.00,0.00,0.00}{\cite{QLi}}, the process of network slicing as a service
within operators by typical examples is illustrated, and the service orchestration and service
level agreement mapping for quality assurance are introduced to illustrate the architecture of
service management across different levels of service models. Network slice instances for specific
purposes are studied as well, for example, the software-defined wireless
networking enabled WLAN slices to provide fine-grained spectrum are introduced in\textcolor[rgb]{1.00,0.00,0.00}{\cite{SDWN}},
and the operator slices to act as mobile virtual network operators are discussed in\textcolor[rgb]{1.00,0.00,0.00}{\cite{SDND2D}}. Meanwhile,
the core network (CN)-based network slicing is presented in\textcolor[rgb]{1.00,0.00,0.00}{\cite{FDNS}}, where two example network
slice instances are illustrated to explain the impact of use case requirements on the network
slice design. In the 5G technical report given by 3rd Generation Partnership Project (3GPP), the
support for network slicing appears as one of the key requirements\textcolor[rgb]{1.00,0.00,0.00}{\cite{NSCN}}.


Despite the evident attractive advantages, the traditional CN-based network slicing also comes
with its own challenges.
First, the conventional network slicing solution is mainly
business driven and only addresses the needs of use cases, which does not highlight the characteristics of radio access networks (RANs) on network slicing creation.
The requested network slicing may not be effective when the radio resource in RANs is in shortage.
As a result, the network slicing should consider the radio transmission impacts, and
the corresponding network slicing jointly considering the status of RANs should be highlighted.
Second, most of the existing work on network slicing is purely based on CNs, while network
slicing as an end-to-end solution should cover the specific characteristics of RANs. Especially in
some cases like uMTC demanding for a ultra low latency, the CN-based slicing with the general
radio access network (RAN) is hard to satisfy with the requirements of 5G performance.

Fog radio access network (F-RAN) has emerged as a promising 5G RAN, which can satisfy
diverse quality of service (QoS) requirements like high spectral efficiency, high energy efficiency, low latency, and
high reliability for different service types in 5G\textcolor[rgb]{1.00,0.00,0.00}{\cite{FRANdiaoyan}}. The traditional CN-based network slicing
cannot take full advantage of the edge characteristics of F-RANs to efficiently support mMTC
and uMTC service types.

To cope with the aforementioned challenges that the CN-based network slicing meets, a
technological framework of a novel network slicing termed by access slicing specified for F-RANs
is proposed in this article. The proposed access slicing is compatible with the existing
CN-based network slicing solution\textcolor[rgb]{1.00,0.00,0.00}{\cite{NSCN}}, and it has a significant difference from both the CN-based
network slicing and the RAN-based access slicing\textcolor[rgb]{1.00,0.00,0.00}{\cite{NSaaS}}. The proposed access slicing can
be regarded to combine the advantages of CN-based and RAN-based network slicing in 5G, and
it can take full advantages of F-RANs. Further, the advanced access slicing is information-aware
to guarantee diverse QoS requirements, and
quality of experience (QoE) requirements as functions of QoS requirements.

In this article, motivated to offer a comprehensive research on the access slicing in F-RANs,
the corresponding architecture is first presented, in which a new management entity (e.g., access
slice orchestration) is defined to fulfill the legacy function of CN-based network slicing. This new
entity helps to orchestrate proper network functions for different access slices and achieves the
coexistence of the multiple access slices. Meanwhile, the resource management between different
slices is exploited, in which the cache and radio resources are jointly designed. Furthermore,
the social-aware slicing technique is presented as well, which assists to create and adjust access
slice instances. In addition, the open issues and future research works are presented as well.

The remainder of this article is organized as follows. The new architecture for the access
slicing in F-RANs will be introduced in Section II. In Section III, key techniques including
resource management and social-aware slicing techniques will be presented. Open issues will be
discussed in Section IV, followed by conclusion in Section V.
\section{Hierarchical Architecture for Access Slicing}

The hierarchical architecture of access slicing in F-RANs is capable to provide numerous
services with the desired QoS, including the typical eMBB, uMTC, and mMTC defined in 5G.
To fulfill the orchestration, control and data functions in networks,
two layers are defined in the proposed access slicing architecture as shown in Fig. \ref{System}: centralized orchestration layer
and slice instance layer. In the centralized orchestration layer,
the access slice orchestration is used to handle the centralized network orchestration for dynamic
provisioning of the slices and manage the resource between the implemented access slice instances.
The slice instance layer consists of various access slice instances that provide the requested
services.
With the guarantee of slice isolation, each access slice instance can operate as a logical separated
network and has specified control/data plane.
Based on the big data analysis and the identified results of the soft-defined access slice orchestration,
the access slicing is implemented to fulfill the edge computing and virtualization capabilities.

\subsection{Key Components}

\begin{figure}
    \centering
    \includegraphics[height=3.5in]{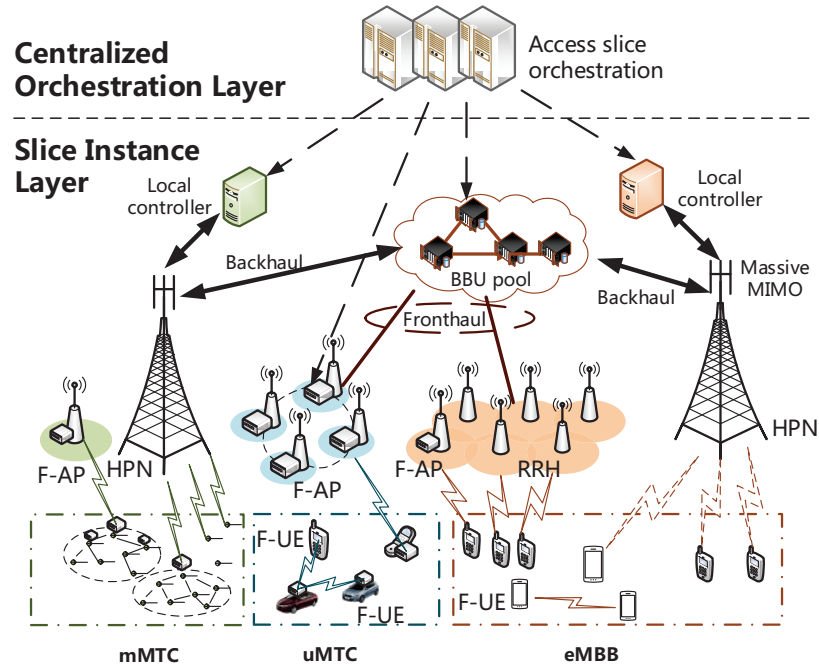}
    \caption{A illustration of access slicing in F-RANs.}
    \label{System}\vspace*{-25pt}
\end{figure}

The proposed access slicing architecture in F-RANs, as illustrated in Fig. \ref{System}, comprises
two kinds of key components: \textit{access slice orchestration} and \textit{access slice instance}.
By analyzing the information collected from the mobile applications,
user equipments, base station, and service platform,
the access slice orchestration identifies the service type. Taking the profiles of existing slice instances into account,
the access slice orchestration determines whether the needed access slice instances should be created.
During the slices' runtime, the access slice orchestration behaves the centralized resource allocation and inter-slice resource management.
Once the access slice instance is in crisis or any accident events happens, the adjustment of slices instance
can be triggered immediately with the help of social-aware mechanisms.

Besides the access slice orchestration, the access slicing architecture includes various instances, such as
mMTC, uMTC, and eMBB, which is described in Fig. \ref{System}. Note that
based on the illustrated eMBB, uMTC, and eMBB instances,
different service types can be handled as a combination of the three service types.
Taking full advantage of the convergence of fog computing, cloud computing, and heterogeneous networking, F-RANs\textcolor[rgb]{1.00,0.00,0.00}{\cite{FRAN}} are able to support different kinds of access slice.
With a substantial amount of storage, communication,
control, configuration, measurement, and management put at the
fog access points (F-APs) and fog user equipments (F-UEs), the
traditional cloud computing paradigm is extended to the network edge. Hence,
both the centralized cloud computing in the baseband unit (BBU) pool and the scalable fog computing composed of F-APs and F-UEs can provide collaboration radio signal processing (CRSP) and cooperative radio resource management (CRRM). The CRSP and CRRM in the BBU pool are centralized and more effective while CRSP and CRRM in the F-APs and F-UEs are closer to the edge and real-time, which makes F-RANs adaptive to the dynamic traffic and time-varied radio environment. Meanwhile, the advanced 5G techniques, such as the massive multiple-input multiple-output (MIMO), non-orthogonal
multiple access, can be directly applied in access slices.

\textbf{\emph{eMBB Instance:}} To support the eMBB service with goals of high data rates and ubiquitous connection, the decomposition of control and data plane
is applied in the eMBB instance. In this instance, high power nodes (HPNs) configured with completed protocol stack functions are mainly used to deliver control signaling and system broadcasting information for accessed UEs with a basic transmit data rate. Massive MIMO technique are configured in HPNs to provide additional diversity and multiplexing gains. In special zones like hot spots, the dense remote radio heads (RRHs) with only radio functions are randomly distributed to provide a high capacity, which are connected to the BBU pool for achieving the large-scale centralized CRSP gains. The BBU pool with physical layer, medium access control layer, and network layer functions is interfaced with HPNs via backhaul,
which indicates that there is coordination between the
BBU pool and the HPN over the backhaul links.
To mitigate the inter-tier interference between F-APs/RRHs and HPNs,
a cooperative processing and scheduling is implemented with the aid of the interface between the BBU pool and HPN.
Through the centralized large-scale CoMP approach, the inter-tier interferences can be coordinated conveniently.

\textbf{\emph{uMTC Instance:}} To meet the stringent latency demand of uMTC service, both the processing delay and transmit delay should be carefully constrained. To decrease the transmit latency, F-APs with cached content are deployed close to the desired UEs. To minimize the processing delay in the air interface, the protocol design in physical layer, medium access control layer, and network layer should be redesigned. A shorten transmission time interval in physical layer is essential to enable a reduced processing delay, as well as a reduced access delay and hybrid
automatic repeat request acknowledgement/nonacknowledgement round trip time. To address the state transition delay issue, the UE radio resource control state should also be optimized. Moreover, the device-to-device communication technique can be used. Since the radio environment is time-varied and the traffic in this instance is delay-sensitive, it is essential to approach the minimal delay via CRRM.

\textbf{\emph{mMTC Instance:}} The realization of mMTC instance is specified with massive connection in a dedicated area, and the clustering mechanism can be considered. As shown in Fig. \ref{System}, the adjacent UEs are formed into a mesh or tree-like topology cluster, wherein the packet traffic generated in the cluster are delivered to the F-AP or HPN via the elective cluster head. To reduce the cost caused by the massive devices, the functions of traditional air interface protocols need to be re-designed elaborately. Considering the small-volume-data and delay-tolerant features, a smaller resource block and a larger transmission time interval in mMTC instances is adopted. The mobile management is simplified and the handover between APs is not supported, since most devices in mMTC scenarios are immobilized. The paging function can be sinked and settled in the local controller of HPN, which avoids the frequent communications between RANs and CNs. To handle the dilemma between massive UEs and scarce resource, the social-aware technique can be used to automatically detect the active UEs, which assists the distributed CRRM to manage the resource flexibly for both intra-clusters and inter-clusters.

\subsection{Hierarchical Architecture}

\begin{figure}
    \centering \vspace*{0pt}
    \includegraphics[height=3in]{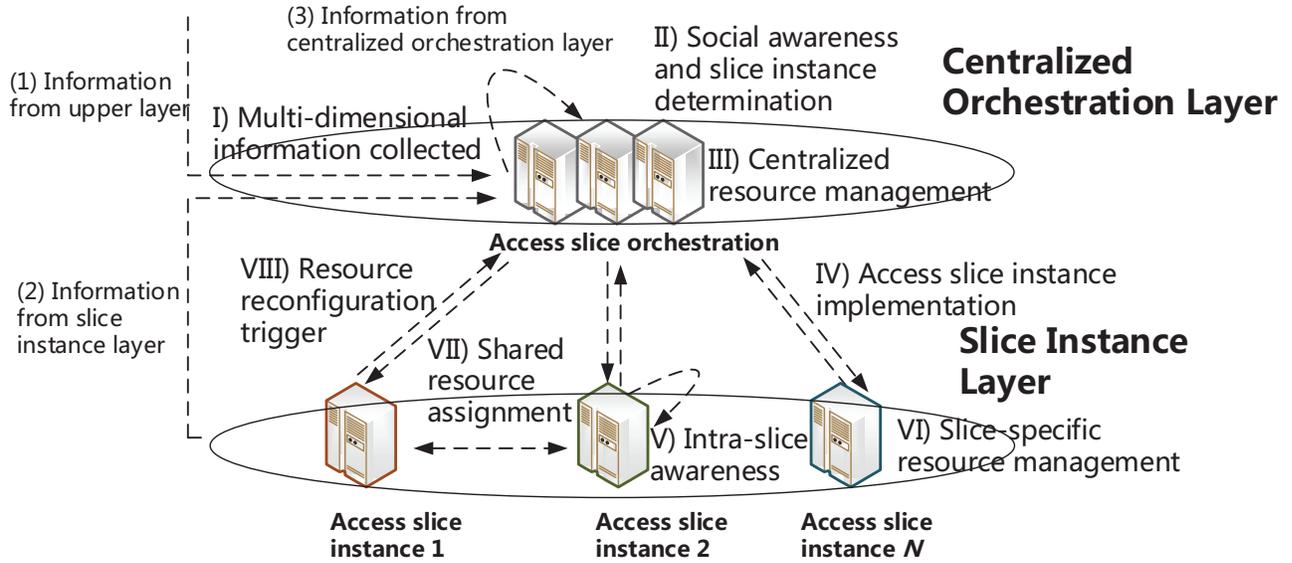}
    \caption{A illustration of the hierarchical architecture.}
    \label{Har}\vspace*{-25pt}
\end{figure}

As shown in Fig. \ref{Har}, the hierarchical architecture of access slicing in F-RANs makes the access slicing adaptively
implement in an convenient way. In the centralized orchestration layer, the multi-dimensional information acquired from devices and nodes are collected, like network, device, application, and others which may influence the provided QoS and QoE. Note that the multi-dimensional information can be divided into: (1) the requested services and subscription information from the upper layer (e.g., the subscriber repository); (2) the characteristics of RANs comprising UE capabilities from the slice instance layer; and (3) the configurations of access slice instances from the centralized orchestration layer. Via the proper machine learning and data mining algorithms, some suggestions or results can be obtained to determine the required access slice instance. It is noted that not only the requirements (e.g. QoS and QoE) of all access slice instances can be fulfilled, but also the performances of the entire F-RAN can be optimized. To optimize the overall F-RAN, the access slice orchestration performs the centralized resources allocation for all slice instances. Thanks to the social-aware technique, both the resource available in the whole F-RAN and the resource needed for each access slice instance can be fully acquired by the access slice orchestration. The access slice orchestration determines the proportion of resources for each slice instance. Note that the resource for each slice instance can be orthogonal or shared, which is determined by the orchestration for the aim to optimize the whole F-RAN.

In the slice instance layer, each access slice instance takes charge of the slice-specific operation and resource management based on the intra-slice awareness.
Due to the introduction of cache featured F-APs and F-UEs, the resource management includes both the radio and the caching resource. The caching resource management has a significant impact on the performance of different slice instances. Like in the case of eMBB instance, F-APs pre-cached the popular content can offload a part of the traffic in hot spots.
On the other hand, the shortage of radio resource has a significant constraint on the performance of slice instances. Hence, it is key to jointly optimize the radio resource and caching resource specified for each access slice instance. Among these different slice instances, some shared resources can be assigned. When the dedicated resource in a instance can not achieve the desired QoS, these shared resource can be used under the permission that the incurring interference is not beyond a pre-defined threshold. If the shared resource has been totally used up, and the performance gap among different slice instances is too big, the resource reconfiguration of all slice instances will be triggered to provide a re-assigned resource allocation solution in the access slice orchestration.

\section{Key Techniques for Access Slicing in F-RANs}

The access slice instances run on the radio hardware and baseband resource pool, which exhibits more elasticity than the pure CN-based network slice solution. However, the guarantee for the completed isolation between slices is challenging in the access slicing. Herein, an enhanced resource management is urgent to enable the slice-specific operation for each isolated access slice. To realize the intelligent control and management of the proposed slicing architecture, the social-aware capability of the network is essential. By realizing the awareness of multi-dimensional information, the access slice instances can be flexibly and optimally designed.

\subsection{Radio and Cache Resource Management}
\begin{figure}
  \centering
  \subfigure[]{
    \includegraphics[width=2.5in]{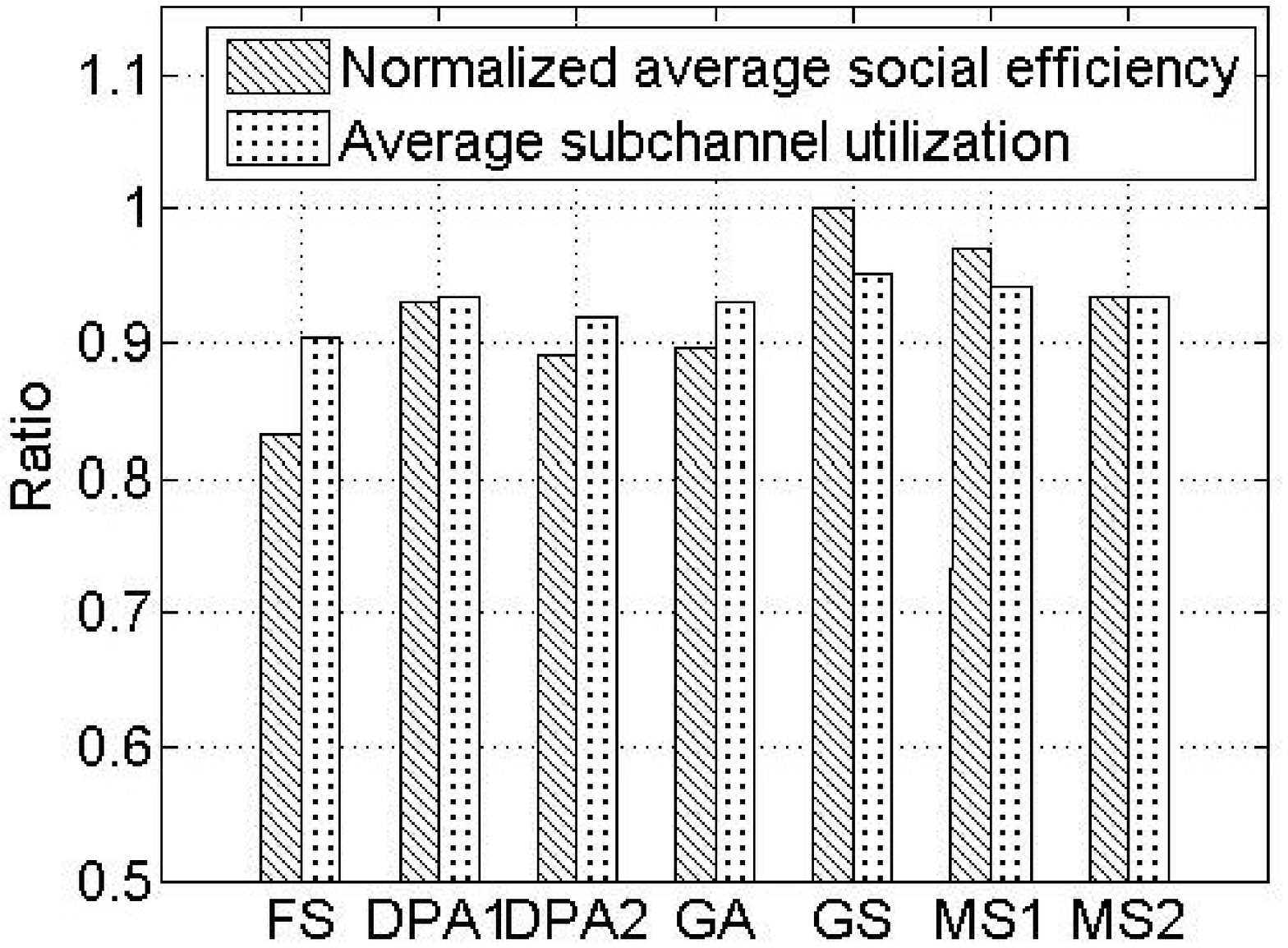}}
  \hspace{1in}
  \subfigure[]{
    \includegraphics[width=2.5in]{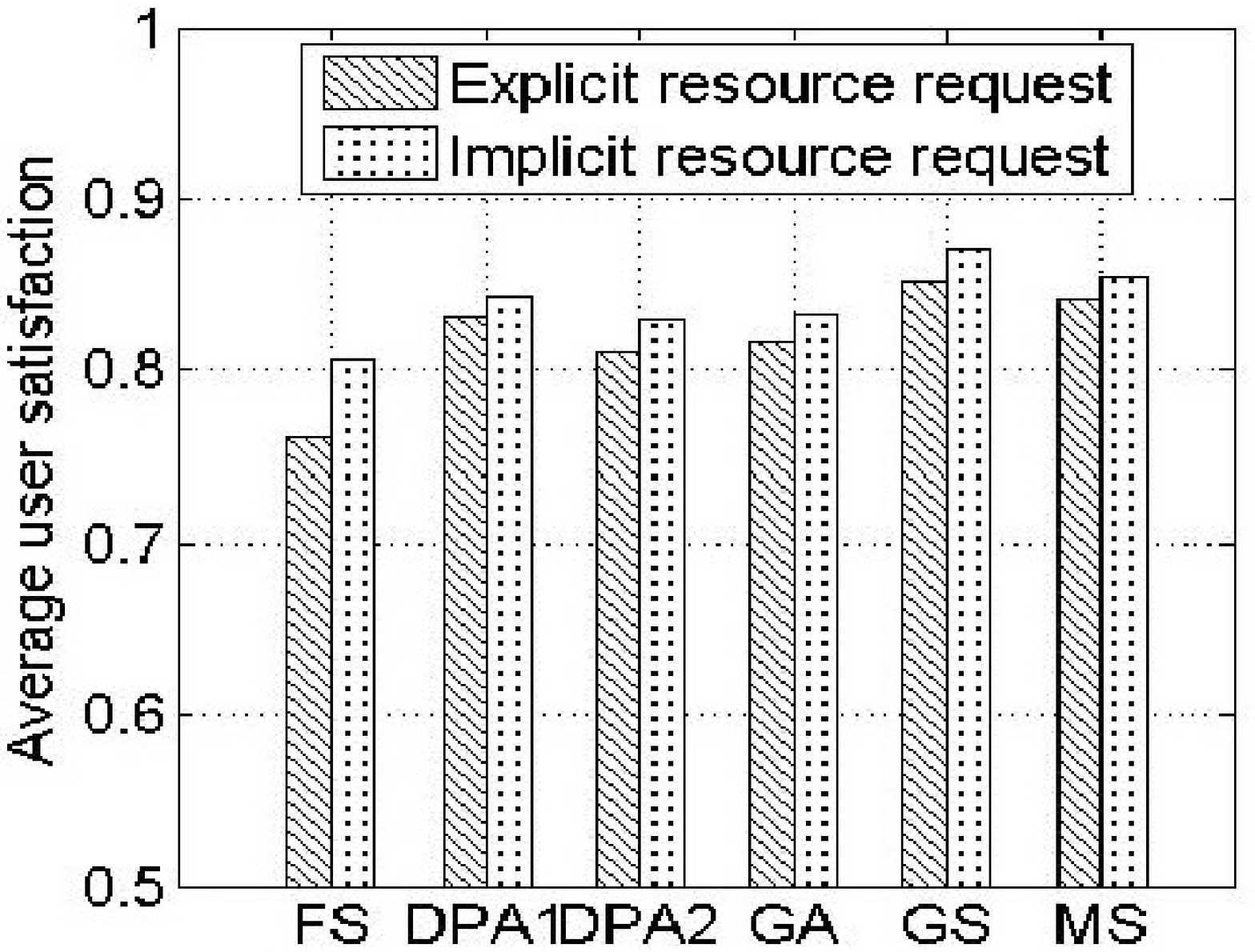}}
  \caption{Performance comparisons among different algorithms applied in the proposed hierarchical auction model\cite{Vir}: a), normalized average social efficiency and average subchannel utilization under different hierarchical auction mechanisms; b) average user satisfaction under different hierarchical auction mechanisms. FS = Fixed sharing, DPA1 = Dynamic programming-based algorithm with group size 1, DPA2 = Dynamic programming-based algorithm with group size 5, GA = Greedy algorithm, GS = General sharing, MS1 = Multiple seller exact solution, MS2 = Multiple seller approximate solution.}
  \label{fig:subfig} 
\end{figure}

An efficient resource management is important to not only accommodate the dynamic
demands of users in slices, but also satisfy the requirements of efficient resource
allocation and isolation between slices.
To enable the intra-slice customization and inter-slice isolation,
the potential mutual influence between slices should be taken into account when behaving the resource management.
In\textcolor[rgb]{1.00,0.00,0.00}{\cite{Vir}}, the resource allocation problem
satisfying the requirements of efficient resource allocation, strict inter-slice isolation, and
intra-slice customization is studied,
in which a hierarchical combinatorial auction mechanism is derived.
Based on the hierarchical auction model,
a winner determination problem is formulated and a sub-efficient semi-distributed resource allocation is proposed.
To evaluate the
performance of the proposed algorithm, average social welfare
(i.e., the sum value of all accepted bids), average
resource utilization (i.e., the proportion of resources utilized),
and average user satisfaction (i.e., the ratio of users
whose resource requests are satisfied) are taken into account. As shown in Fig. \ref{fig:subfig}. a and Fig. \ref{fig:subfig}. b,
the proposed algorithm can achieve a relative higher average social welfare, average
resource utilization, and average user satisfaction than the other baselines,
including the fixed sharing scheme where the mobile virtual network operators are preassigned a
fixed subset of resources.

For the access slicing in F-RANs,
the joint resource optimization of radio resource and cache resource is often complex due to the
tight relationship between radio and caching resources. F-APs with larger cache capacity are capable
of caching more requested files and performing a more sophisticated cooperation between them, so that the radio resource will be
reduced. On the other hand, if a AP is assigned more radio
resources, the less files need to be stored, thus saving the caching resources.
To address the difference caused by the cache as a kind of resource,
a framework for edge
cache based performance analysis and radio resource allocation is given in\textcolor[rgb]{1.00,0.00,0.00}{\cite{FRANdiaoyan}},
where a comprehensive summarization on the recent advances of the two issues in F-RANs is presented.
In particular, the optimization on the spectral efficiency, energy efficiency, and latency via resource allocation is quite different from that in existing works for the traditional wireless networks. The cell association to optimize spectral efficiency needs to consider the scenario that users prefer to access F-APs when the contents they need have been locally cached. As a result, the serving F-AP cluster is determined by both the F-AP caching contents and the reference signal received power. Similarly, the energy efficiency optimization should highlight the effect of local caching because the energy consumption in the local F-AP increases but with the benefit that the energy consumption of the backhaul is decreased. Besides, latency optimization problem is becoming challenging due to the coexistence of various transmission modes in the F-RAN. The BBU pool can store numerous contents but only provide a latency-tolerant service, while F-APs with limited cache volume can provide contents with a small latency. Hence, the tradeoff should be balanced between the BBU pool and F-APs with the joint consideration of spectral and energy efficiency.
In addition to the energy efficiency, spectral efficiency, and latency, there are also other new performance metrics taken into account when behaving resource management. In \textcolor[rgb]{1.00,0.00,0.00}{\cite{recentwork}}, the serviceability is proposed
as a key foundational criteria satisfying the IoT paradigm in the 5G era.
The serviceability is defined as the ability of a network to serve UEs within desired requirements (e.g., throughput, delay, and packet loss).
Given the distribution of cached content and the desired data rate constraints, a serviceability maximization problem in F-RAN is proposed, which is handled with an adaptive resource balancing scheme.

\subsection{Social-Aware Slicing Techniques in F-RANs}

\begin{figure}
  \centering
  \subfigure[]{
    \includegraphics[width=2.5in]{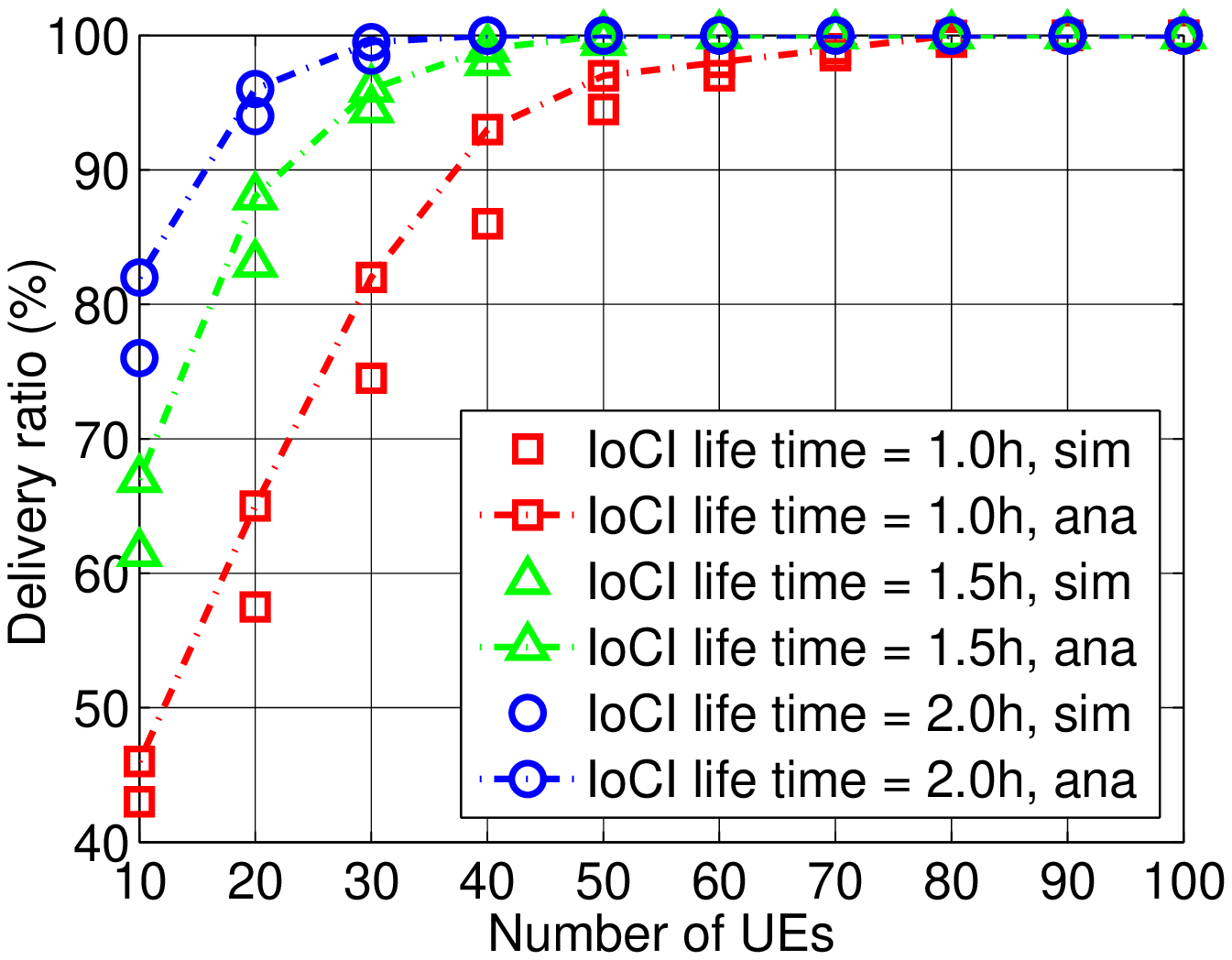}}
  \hspace{1in}
  \subfigure[]{
    \includegraphics[width=2.5in]{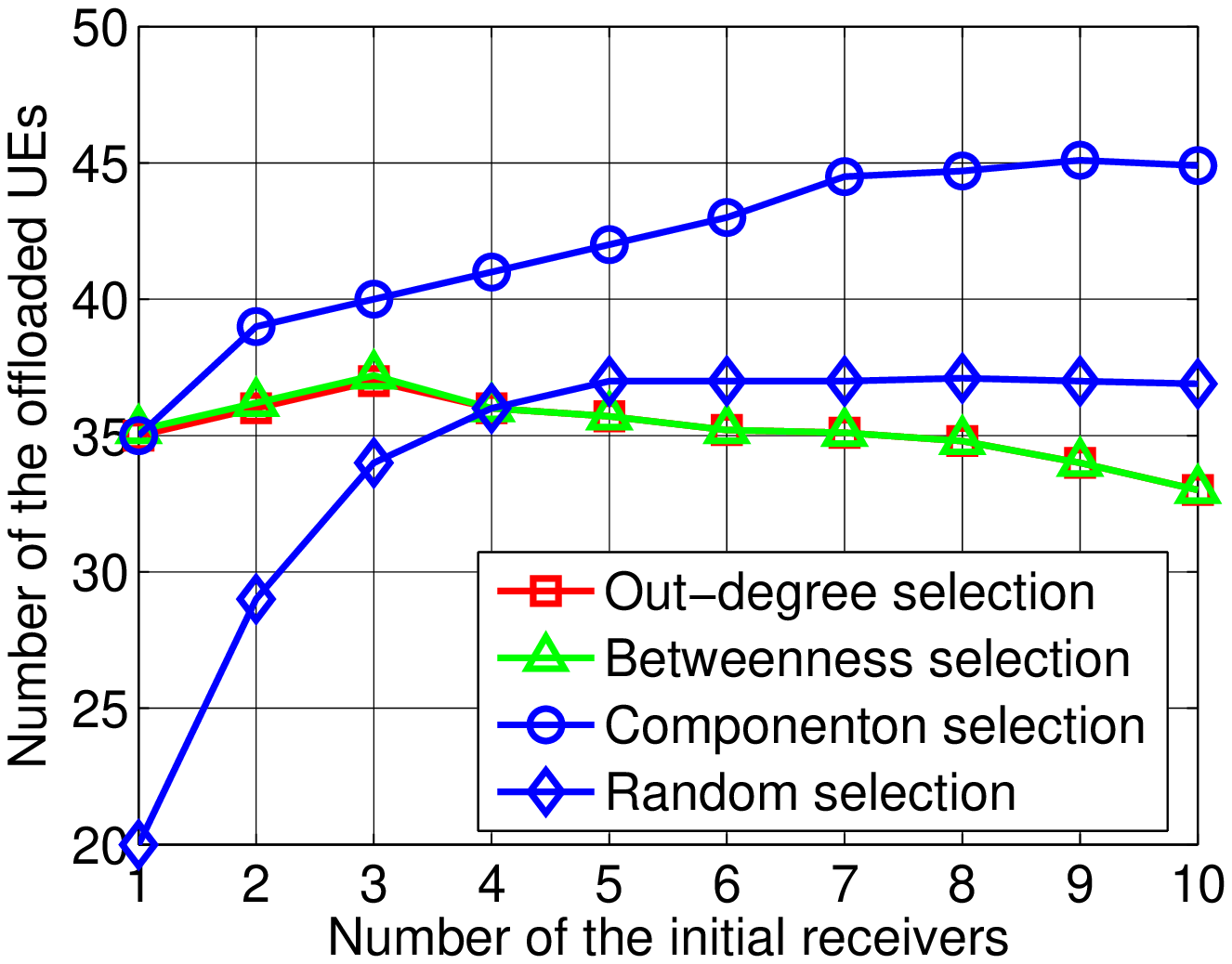}}
  \caption{Performance results for the extending coverage and offloading traffic scenarios in which opportunistic communication are applied\cite{Soc}: a), the average successful delivery ratio before the expiry of IoCI; b), the average number of UEs receiving the IoCI under different selection scheme of the initial
    receiver set.}
  \label{SocWhole} 
\end{figure}

The social awareness is an emerging approach to enable the realization of network slices
in a convenient way. As more and more communications occur between UEs with close relationships,
the network is getting increasingly human-centric and social-aware.
To exploit UEs' behaviors and interactions in the social domain, social awareness turns out to be a
indispensable information for the design and optimization of RANs. By exploiting social properties of nodes and UEs in RANs, social awareness
helps to provide an efficient resource allocation and networking. Inspired by the attractive advantages of social awareness, a detailed survey on the cross-disciplinary research area of social network analysis aided telecommunication
networking is shown in\textcolor[rgb]{1.00,0.00,0.00}{\cite{Soc}}.
In particular,
two eMBB-specific application scenarios are studied to illustrate the benefits of the social awareness:
the extending coverage scenario and the offloading traffic scenario.
In the extending coverage scenario, the limited base stations are sparsely distributed in a large area.
To extend the coverage, the opportunistic communications
amongst the roaming UEs are exploited based on opportunistic contacts in social aware domain.
The roaming UEs with the information of common interest (IoCI) carried can
deliver the IoCI to the other UEs via the opportunistic links once they meet.
It is demonstrated in Fig. \ref{SocWhole}. a that with the aid of opportunistic communication, the delivery ratio of the
IoCI in the extending coverage scenario increases from 45 percentages to 100 percentages. Moreover, the delivery ratio is improved furthermore
as the number of UEs increases and the IoCI life time becomes longer, since
more UEs participate in the information dissemination
process and the UEs can tolerate a longer latency of receiving the IoCI.


In the offloading traffic scenario, numerous UEs are densely crowded in a small area occasionally.
To alleviate the burden of base stations and improve the spectrum efficiency,
the large-scale opportunistic networks are utilized to
offload some traffic from cellular networks.
Based on the social contact graph of the UEs and specific selection schemes, the initial receiver set is determined
and a opportunistic network is constructed.
The BSs firstly inject some copies of the IoCI into the
opportunistic network. Thereafter, the IoCI is disseminated
via opportunistic contacts of UEs as well as via direct injections
from the BSs.
With the aid of opportunistic multicast scheduling in the small-scale
opportunistic scenario, the opportunistic communication between UEs can be established and the heavy traffic is offloaded.
To validate the benefits of opportunistic communication applied
in the offloading traffic scenario, the contact traces of 78 UEs in a crowed area are studied based on the realistic mobility trace of
Infocom 2006. As shown in Fig. \ref{SocWhole}. b, the opportunistic communication is capable of offloading
as high as 58 percentages of the traffic (i.e., 45 UEs) from the cellular networks.

\section{Challenging Work and Open Issues}

Although the access slicing in F-RANs has been initially researched and some primeval
progresses have been achieved, there are still many challenges and open issues remain to be
discussed in the future. Besides the key techniques initially researched in the aforementioned section, the standard development and field trials are other two important issues.

\subsection{Access Slicing Standardization}

Given the relative infancy, the current standardization efforts on network slicing are still in the early stage.
Lots of industry-wide studies about network slicing are on-going and a large number of related organizations are involved, including the Next Generation Mobile Networks (NGMN) Alliance, 3GPP,
and 5G Public Private Partnership (5GPPP).
In NGMN, a detailed definition and application of network slicing has been output. The network slice composed business application layer
is defined as an indispensable part of the 5G architecture\textcolor[rgb]{1.00,0.00,0.00}{\cite{2}}.
In 3GPP, network slicing is identified as one of the key
technologies to be developed in 5G standardization. The study on the CN-based network slicing has been undertaken in System Architecture 2 working group\textcolor[rgb]{1.00,0.00,0.00}{\cite{NSCN}}, where five work tasks for network slicing
have been defined and the corresponding solutions like network slice selection and identification have been discussed.
The realization of network slicing in RANs has been incorporated into the future work schedule. In particular, the key principles for the support of network slicing in RANs have been identified, while the issue about the resource management and isolation among slices are still under investigation.
In 5GPPP, several projects related to the network slicing have been activated, and the architecture enablers and resource management for network slicing related to diverse 5G services are listed in\textcolor[rgb]{1.00,0.00,0.00}{\cite{5GPPP}}. However, the current discussions on network slicing in the industry are mainly focused on the concept and requirements of
network slicing, the solutions for the realization of network slice, and the potential CN-based slicing architecture. It is anticipated that more attention will be focused on the access slicing since it combines the advantages of both CN-based and RAN-based slicing.

Besides, there are a large number of research projects on the development of F-RANs. Numerous works have been done for the related fog computing, as well as other similar concepts like the mobile edge computing and cloudlet. As illustrated in Table. \ref{Compare}, the concepts of these similar concepts are partially overlapping and complementary. All of these similar concepts are driven by an anticipated future characterized by IoT, which can be categorized into the mMTC with low data rate and minimal power consumption and the uMTC with a ultra-low latency. Unlike mobile edge computing and cloudlet,
which use dedicated servers and devices, F-RAN
can utilize any device with cache and computing capabilities at anywhere in the network to enhance the system performance.
Thus the F-RAN offers more flexibility in the choice of devices and can be implemented in a shorter time. Moreover, the Open Fog Consortium has established the foundation for an open fog computing reference architecture. As a result, the access slicing in F-RANs has a promising development, and its standardization should be exploited, which can be regarded as a further step in the existing network slicing standards in various standard organizations.

\begin{table}[h]
\begin{tabular}{|p{2cm}|p{4cm}|p{4cm}|p{4cm}|}
\hline
\multicolumn{1}{|c|}{\textbf{}} & \multicolumn{1}{|c|}{\textbf{Fog computing}}& \multicolumn{1}{|c|}{\textbf{Mobile edge computing}}
&  \multicolumn{1}{|c|}{\textbf{Cloudlet}}  \\
\hline
{\textbf{Developed by}}
& Cisco in 2011
& ETSI as an industry specification in 2014
& Carnegie Mellon University in 2013, later supported by various companies including Intel, Huawei, Vodafone\\
\hline
{\textbf{Concepts}}
& An extension of the cloud computing in the edge of networks
& Push cloud computing capabilities close to the RAN in 4G, 5G
& The middle tier of a 3-tier hierarchy: `mobile device-cloudlet-cloud' \\
\hline
{\textbf{Features}}
& i) A completely distributed, multi-layer network architecture, ii) any device with cache and computing capabilities can be incorporated into the network and iii) achieving a cooperation between computing, communication and control is characterized
& A number of standardized MEC servers and applications are deployed at the edge of network, which is capable of providing the cache, analysis, processing and control functions
& i) Extra management is no need, ii) the implementation is based on virtual machines and Openstack++ platform, and iii) a fairly strong computing ability is provided\\
\hline
{\textbf{Main standardization actors}}
& Open Fog & ETSI & Open Edge Computing\\
\hline
{\textbf{Typical use cases}}
& IoT, optimized local content distribution, data caching, and so on.
& Video analytics, location services, IoT, augmented reality, and so on
& Face detection, voice recognition (used for traffic offloading), emergency scenario, and so on\\
\hline
\end{tabular}
\caption{Comparison of fog computing, mobile edge computing, and cloudlet.}
\label{Compare}
\end{table}

\subsection{Testbed and Field Trials}

The industry has made lots of efforts on the realization and test of CN-based network slicing.
During the Mobile World Congress 2016, Deutsche Telekom and Huawei demonstrated the
world’s first end-to-end 5G system to validate the CN-based network slicing for diverse 5G use
cases, in which the autonomous end-to-end network slicing adding the dynamic and real-time
slicing of the 5G RAN, CN, and the interconnecting transmission network has been implemented
and the corresponding results have been shown. These demonstrations have shown that the CN-based
network slices can be automatically created in an optimized way with a cost of sub-minute
time. Motivated by the CN-based network slicing testbed, it can be anticipated that the access
slicing in F-RANs can be tested in the future 5G field trials, and the gains will be significant.

To show the performance gains achieved by the proposed access slicing, the first step is
to build the powerful F-RAN testbed. Then, the proposed architecture and key techniques for
access slicing should be fulfilled in the implemented F-RAN testbed. Notable achievements will
be anticipated to be gained and preliminary results for the access slicing in trial tests will be
output.

\section{Conclusion}

In this article, an access slicing in fog radio access networks (F-RANs) has been proposed.
Compared to the separated CN-based and RAN-based network slicing, the proposed slicing
combines the advantages of the CN-based and RAN-based network slicing, which taking full
advantages of the edge characterizes of F-RANs. To enable the implementation of the proposed
access slicing architecture, the body of key techniques are broadly divided into hierarchical
resource management and multi-dimensional social-aware slicing. With the goal of understanding
further intricacies of key techniques, diverse problems within these key techniques have been
summarized, as well as corresponding solutions that have been presented. Nevertheless, given
the relative infancy, there are still quite a number of outstanding problems that need further
investigation. Notably, it is anticipated that great attention will be focused on the progress
of standardization development and trial tests on access slicing in F-RANs, which makes the
commercial rollout of access slicing as early as possible.





\begin{IEEEbiography}[{\includegraphics[height=1.25in]{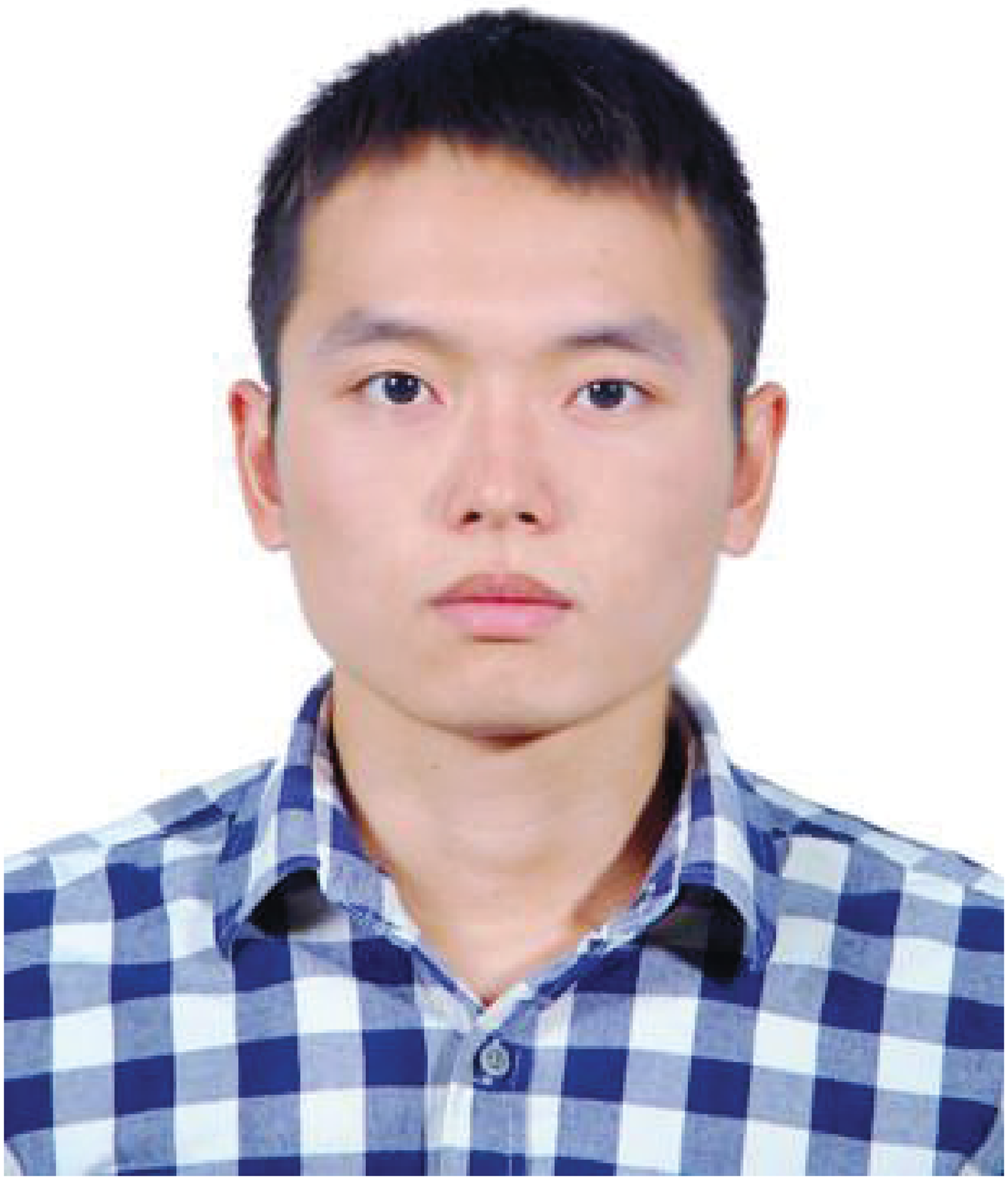}}] {Hongyu
Xiang} (xianghongyu88@163.com) received the B.E. degree in
telecommunication engineering from Fudan University, China, in 2013. He is currently
pursuing the Ph.D. degree at the key laboratory of universal wireless
communications (Ministry of Education) at BUPT.
His research focuses on cooperative radio resource management and collaborative
radio signal processing in heterogeneous
cloud radio access networks (H-CRANs) and
fog radio access networks (F-RANs).
\end{IEEEbiography}

\begin{IEEEbiography}[{\includegraphics[height=1.25in]{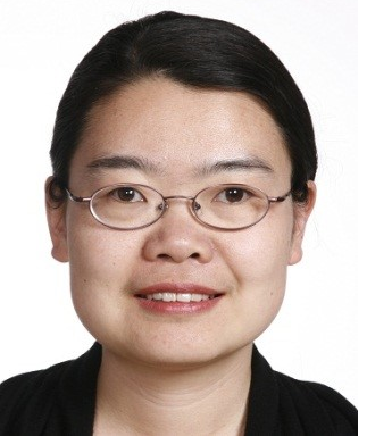}}]
{Wenan Zhou} (zhouwa@bupt.edu.cn) is currently an associate Professor of School of Computer Science, BUPT. She received her Ph.D. degree in electrical engineering from Beijing University of Posts and Telecommunications (BUPT) in 2002. In 2007, she furthered her study of broadband wireless communication technology in the University of California, San Diego (UCSD) as a Visiting Scholar. Her current research interests include wireless mobile communication theory, Radio Resource Management and QoE management in 5G.
\end{IEEEbiography}

\begin{IEEEbiography}[{\includegraphics[height=1.25in]{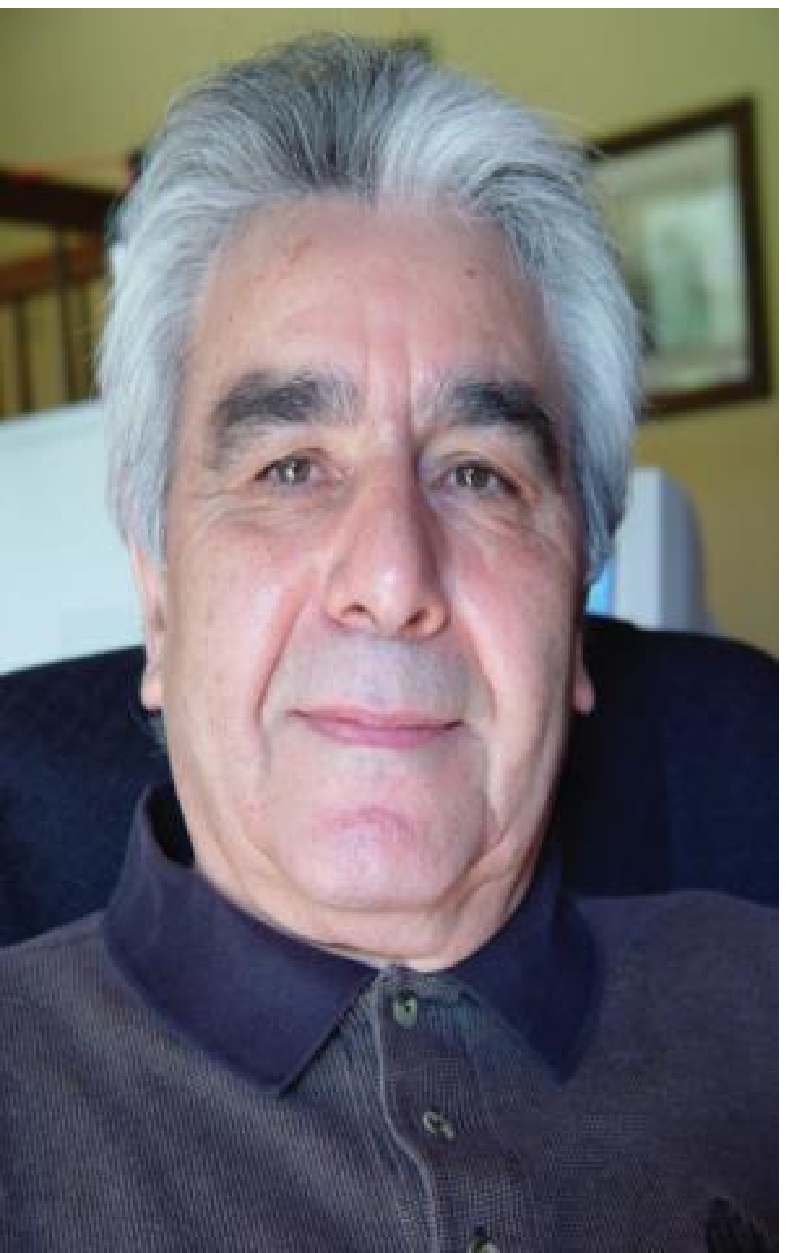}}]
{Mahmoud Daneshmand} (mdaneshm@stevens.edu)) received his Ph.D. and M.S. degrees in
statistics from the University of California, Berkeley, USA. He is
currently a professor in the School of Business at Stevens Institute
of Technology. He is an expert in big data analytics, data mining algorithms, machine
learning, probability and stochastic processes, and statistics. He
is a co-founder and chair of the Steering Committee
of the IEEE IoT Journal, and IEEE Transactions on Big Data.
\end{IEEEbiography}

\begin{IEEEbiography}[{\includegraphics[width=1.0in]{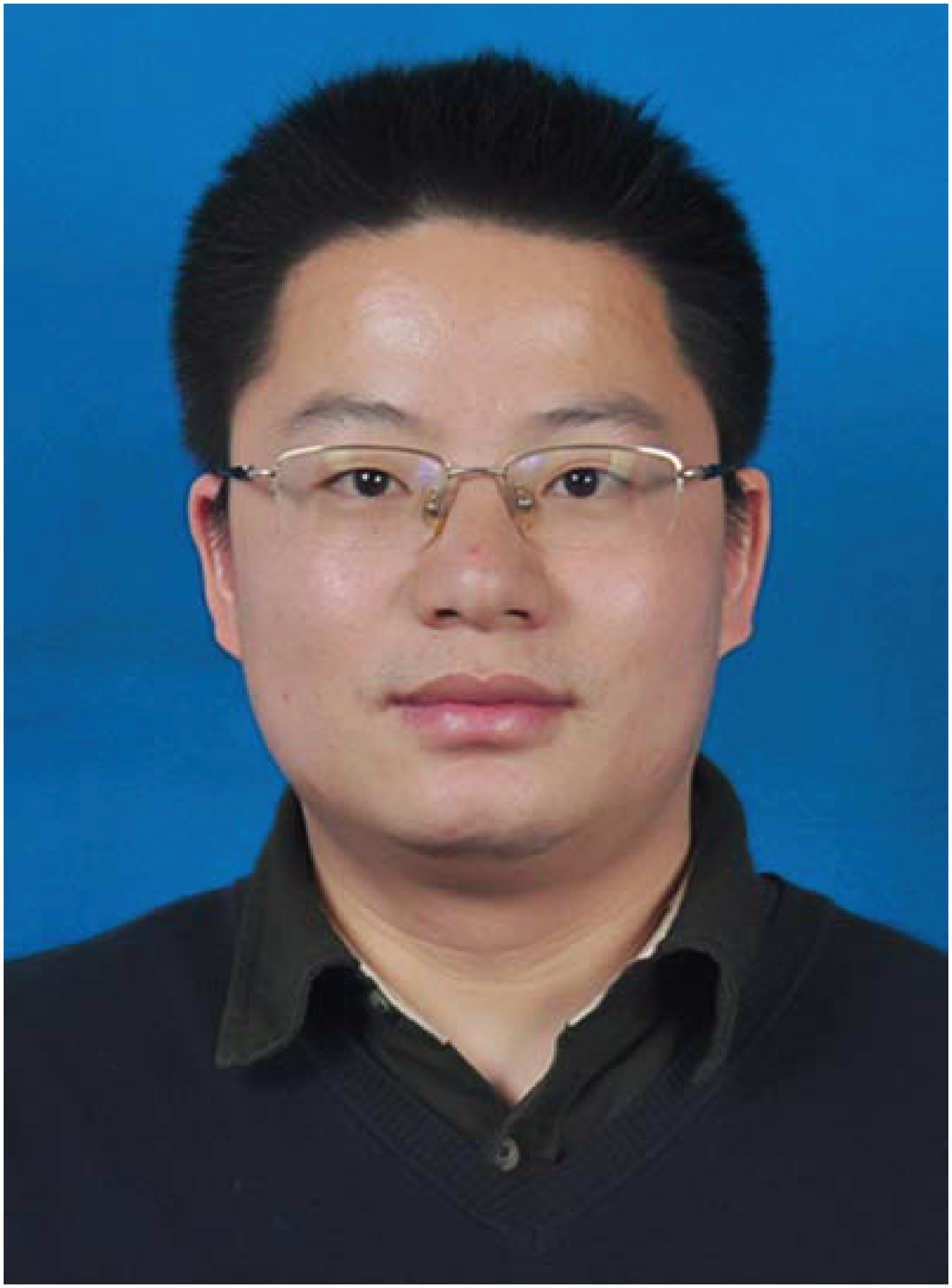}}]
{Mugen Peng}[M'05--SM'11] (pmg@bupt.edu.cn) received the Ph.D. degree from the Beijing University of Posts and Telecommunications (BUPT), China, in 2005. He has a Full Professor in BUPT since 2012. His main research interests focus on cooperative communication, self-organization networking, heterogeneous networking, cloud communication, and internet of things. He was a recipient of the 2014 IEEE ComSoc AP Outstanding Young Researcher Award, and the best paper award in JCN, IEEE WCNC 2015, WASA 2015, GameNets 2014 and so on.
\end{IEEEbiography}

\end{document}